# Acoustic THz graphene plasmons revealed by photocurrent nanoscopy

*Pablo Alonso-González[1,2†], Alexey Y. Nikitin[1,3*], Yuanda Gao[4*], Achim Woessner[5*], Mark B. Lundeberg[5], Alessandro Principi[6], Nicolò Forcellini[7], Wenjing Yan[1], Saül Vélez[1], Andreas. J. Huber[8], Kenji Watanabe[9], Takashi Taniguchi[9], Félix Casanova[1,3], Luis E. Hueso[1,3], Marco Polini[10], James Hone[4], Frank H. L. Koppens[5,11], and Rainer Hillenbrand[3,12†]*

[1]CIC nanoGUNE, E-20018, Donostia-San Sebastián, Spain
[2]Departamento de Física, Universidad de Oviedo, Oviedo 33007, Spain
[3]IKERBASQUE, Basque Foundation for Science, 48011 Bilbao, Spain
[4]Department of Mechanical Engineering, Columbia University, New York, NY 10027, USA
[5]ICFO-Institut de Ciencies Fotoniques, The Barcelona Institute of Science and Technology, 08860 Castelldefels (Barcelona), Spain
[6]Radboud University, Institute for Molecules and Materials, NL-6525 AJ Nijmegen, The Netherlands
[7]Department of Physics, Imperial College London, London SW7 2AZ, United Kingdom
[8]Neaspec GmbH, Martinsried, Germany
[9]National Institute for Materials Science, 1-1 Namiki, Tsukuba 305-0044, Japan
[10]Istituto Italiano di Tecnologia, Graphene labs, Via Morego 30 I-16163 Genova, Italy
[11]ICREA – Institució Catalana de Recerca i Estudis Avancats, Barcelona, Spain
[12]CIC NanoGUNE and EHU/UPV, E-20018, Donostia-San Sebastian, Spain

* These authors contributed equally
†Correspondence to: r.hillenbrand@nanogune.eu, p.alonso@nanogune.eu

**Terahertz (THz) fields are widely applied for sensing, communication and quality control[1]. In future applications, they could be efficiently confined, enhanced and manipulated – well below the classical diffraction limit - through the excitation of graphene plasmons (GPs)[2,3]. These possibilities emerge from the strongly reduced GP wavelength, $\lambda_p$, compared to the photon wavelength, $\lambda_0$, which can be controlled by modulating the carrier density of graphene via electrical gating[4,5,6,7,8]. Recently, GPs in a graphene-insulator-metal configuration have been predicted to exhibit a linear dispersion (thus called acoustic plasmons) and a further reduced wavelength, implying an improved field confinement[9,10,11], analogous to plasmons in two-dimensional electron gases (2DEGs) near conductive substrates[12]. While infrared GPs have been visualized by scattering-type scanning near-field optical microscopy (s-SNOM)[6,7], the real-space imaging**

**of strongly confined THz plasmons in graphene and 2DEGs has been elusive so far - only GPs with nearly free-space wavelength have been observed[13]. Here we demonstrate real-space imaging of acoustic THz plasmons in a graphene photodetector with split-gate architecture. To that end, we introduce nanoscale-resolved THz photocurrent near-field microscopy, where near-field excited GPs are detected thermoelectrically[14] rather than optically[6,7]. The on-chip GP detection simplifies GP imaging, as sophisticated s-SNOM detection schemes can be avoided. The photocurrent images reveal strongly reduced GP wavelengths ($\lambda_p \approx \lambda_0/66$), a linear dispersion resulting from the coupling of GPs with the metal gate below the graphene, and that plasmon damping at positive carrier densities is dominated by Coulomb impurity scattering. Acoustic GPs could thus strongly benefit the development of deep subwavelength-scale THz devices.**

The graphene photodetector is illustrated in Fig. 1a. A monolayer graphene sheet was encapsulated between two h-BN layers[15]. The h-BN(13 nm)-graphene-h-BN(42 nm) heterostructure is placed on top of a pair of 15-nm-thick AuPd gates, which are laterally separated by a gap of 50 nm. Applying individual voltages to the gates allows for controlling independently the carrier concentrations $n_1$ and $n_2$ in the graphene sheet at the left and right sides of the gap.

In Fig. 1a we also introduce the concept of THz photocurrent nanoscopy, and its application for GPs mapping. The setup is based on a s-SNOM (Neaspec), where the metal tip is illuminated with the THz beam of a gas laser (SIFIR-50 from Coherent, providing output power in the range of a few 10 mW). Owing to a lightning-rod effect, the incident field is concentrated at the tip apex yielding a THz nanofocus[16]. Once brought into close proximity of the sample, the near fields of the nanofocus induce a current in the graphene sheet, similar to IR photocurrent nanoscopy[14,17]. Recording the current as a function of the tip position yields nanoscale-resolved THz photocurrent images. For the current measurement, the graphene is contacted electrically in a lateral geometry (i.e. metal contacts were fabricated at both sides of the heterostructure, as shown in Fig. 1a). Analogously to s-SNOM[18] and scanning photocurrent nanoscopy[14,17], we isolate the near-field contribution to the total photocurrent, $I_{PC}$, by (i) oscillating the tip vertically at frequency $\Omega$ and (ii) demodulating the detector signal at $2\Omega$. This technical procedure is required because of the background photocurrent generated by the diffraction limited illumination spot. We achieved a spatial resolution of about 50 nm (supplementary information S1), which is an improvement of nearly 4 orders of magnitude compared to diffraction-limited THz imaging.

Fig. 1b shows a photocurrent image of the photodetector, recorded at 2.52 THz ($\lambda_0$ = 118.8 μm). Choosing graphene charge carrier densities $n_1$ = 0.77 and $n_2$ = -0.77x10$^{12}$ cm$^{-2}$, we generate a sharp pn-junction in the graphene above the gap between the gates. We observe a strong near-field photocurrent, $I_{PC}$, which is localized to an about 1 μm wide region centred above the gap (central part of Fig. 1b). It can be explained by a photo-thermoelectric effect: due to a variation of the local Seebeck coefficient $S$ in graphene (generated by the carrier density gradient), a local temperature gradient (caused by the THz nanofocus at the tip apex) generates a net charge current[14, 17]. Because the variation of the carrier concentration – and thus ΔS – is largest between the two gates, we expect a maximum in the photocurrent at this location. In Fig. 1b, however, we observe a slight decrease of the photocurrent between the gates. We explain it by the reduced near-field intensity when the tip is above the gap, owing to the weaker near-field coupling between the tip and the metal gates. To corroborate the photo-thermoelectric origin of the THz photocurrent, we carried out local near-field photocurrent measurements at the gap location as a function of $n_1$ and $n_2$. The obtained near-field photocurrent pattern, $I_{PC}$ ($n_1$, $n_2$), exhibits six regions of alternating signs (supplementary information S2), which is a characteristic signature of the thermoelectric effect[19, 20].

Intriguingly, near the pn-junction we observe photocurrent oscillations perpendicular to the graphene edge (indicated by the horizontal white line in the upper part of Fig. 1b), which decay with increasing distance from the edge (supplementary information S3). Resembling the s-SNOM images of infrared GPs[6, 7], we attribute them to THz GPs - collective excitations of 2D mass-less electrons coupled to THz fields. The mechanism of the photocurrent generation is studied in detail in ref.[14], which we summarize here: The near fields at the tip apex launch radially propagating GPs, which interfere with their own reflections from the graphene edge, producing oscillations of the electric field intensity – and thus the local energy dissipation – when the tip is scanned perpendicular to the graphene edge. The dissipated energy heats the pn-junction, yielding a photo-thermally induced current that oscillates with a period of half the plasmon wavelength $\lambda_p/2$. We corroborate the plasmonic origin of the photocurrent oscillations by recording line profiles along the dashed black line in Fig. 1b at different illumination frequencies but fixed carrier densities $n_1$ = 0.77x10$^{12}$ cm$^{-2}$ and $n_2$ = 1.11x10$^{12}$ cm$^{-2}$. We find that the oscillation period decreases with increasing illumination frequency $f$ (Fig. 2a). By measuring the oscillation period as a function of $f$ (see supplementary information S4), we obtain the dispersion relation ($f$ vs. Re(q$_p$) = $2\pi/\lambda_p$, with q$_p$ being the complex-valued GP wavevector), which is shown in Fig. 2b (red symbols). Interestingly, we find a nearly linear dispersion at low frequencies, in excellent agreement with the calculated GPs dispersion (blue contour plot in Fig. 2b). Note that the calculations (see Methods) take into account the different layers of the heterostructure, as well as the metal gates (air/h-BN/G/h-BN/AuPd/SiO$_2$). The linear dispersion is typical for acoustic plasmons[9, 10, 11, 21] but in strong contrast to conventional GPs in free-standing graphene, where $f \propto \sqrt{Re(q_p)}$ (blue solid curve in Fig. 2b)[2, 3]. We conclude that acoustic GPs rather than conventional GPs are observed.

Interestingly, the GP wavelengths in the heterostructure are reduced by a factor of 12 compared to GPs of free-standing graphene, i.e. a factor of about 70 compared to $\lambda_0$. We further highlight the small slope of the GP dispersion. It corresponds to a group velocity $v_g \approx$ 0.014 c, which is about one order of magnitude smaller than that of GPs in free-standing graphene at $f$ = 2.52 THz. For low frequencies, the group velocity can be approximated by the analytical expression in Eq. (1). It is derived within a random phase approximation approach, where for the effective electron-electron interaction in graphene we took into account screening stemming from both h-BN slabs and metal gate[11] (supplementary information S5 and S6):

$$v_g = v_F \frac{\epsilon_z + 8d\frac{k_F e^2}{4\pi\varepsilon_0 \hbar v_F}}{\sqrt{\epsilon_z\left(\epsilon_z + 16d\frac{k_F e^2}{4\pi\varepsilon_0 \hbar v_F}\right)}} \qquad (1)$$

$\epsilon_z$ is the out-of-plane h-BN permittivity, $k_F = \sqrt{\pi n_2}$ is the Fermi wave number, $v_F = 10^6$ m/s is the graphene Fermi velocity, and $d$ is the bottom h-BN thickness. The result of Eq. (1) for our heterostructure is displayed in Fig. 2b by the dashed black line. At low frequencies, we observe an excellent agreement with the experimental (symbols) and calculated (blue color plot) dispersion. Eq. (1) further predicts a decreasing group velocity with decreasing spacing *d* between the metal and the graphene, which will be subject of future studies.

The strongly reduced wavelength of the acoustic THz GPs implies an extreme plasmonic field confinement, which we study by numerical electromagnetic simulations (see Methods) of the real part of the GPs´ vertical near-field distribution, $Re(E_z(x,z))$ (Figs. 3a,b). To that end, we place a dipole source above the h-BN encapsulated graphene on gold (inset Fig. 3b). For comparison, we place a dipole source above a free-standing graphene sheet to excite conventional GPs (inset Fig. 3a). For the free-standing graphene (Fig. 3a) we observe propagating GPs with an anti-symmetric near-field distribution[22, 23], where the out-of-plane field decay length is $\lambda_p/2\pi$. In strong contrast, for the GPs in the heterostructure (Fig. 3b) we observe an asymmetric near-field distribution, where the GP field is concentrated inside the 42 nm-thick h-BN layer between the metal and the graphene (see zoom-in of Fig. 3b). This deep-subwavelength-scale vertical (z-direction) confinement of about $\lambda_0/2800$ cannot be achieved by pure dielectric loading (see supplementary information S7). Inside the h-BN, the field $E_z$ is constant across the layer, owing to the anti-symmetric distribution of charge carriers (illustrated in the zoom-in of Fig 3b). We explain this finding by the hybridization of GPs in the graphene sheet with their mirror image in the AuPd layer (illustrated in Fig 3d), yielding an anti-symmetric (short wavelength) GP mode, analogous to out-of-phase plasmon modes in double layer graphene[24, 25]. The out-of-phase charge oscillation between the graphene and the gold surface confirms that indeed an acoustic GP mode is observed.

To quantify the near-field distribution, we show in Fig.3c simulated near-field profiles of $|Re(E_z)|$ (extracted along the dashed lines in Figs. 3a, and 3b). We observe that the near field inside the bottom h-BN layer is about 36 times larger than the near field at the surface of the free-standing graphene. A strong near-field enhancement is also observed at the surface of the top h-BN layer, which is 5 times larger than on the free-standing graphene. For the out-of-plane decay length we find $\delta_1 = \frac{\lambda_p}{2\pi} = 3.23\ \mu m$ for free-standing graphene. It is reduced to $\delta_2 = 0.26\ \mu m$ for the heterostructure (Fig. 3c), owing to the short wavelength of the acoustic GPs.

We also studied the plasmon interference pattern (along the dashed black line in Fig. 1b) as a function of charge carrier density $n_2$ (Fig. 4a). We observe that the fringe spacing (plasmon wavelength) increases with increasing carrier concentration of both electrons and holes, demonstrating that acoustic GPs can be tuned by electrical gating, similar to plasmons in single layer graphene [4, 6, 7]. However, in strong contrast to IR GPs, we observe THz GPs at even the charge neutrality point (simultaneously probed by direct-current electrical measurements and indicated by a dotted white line in Fig. 4a). The photocurrent profile at the charge neutrality point (Fig. 4b) clearly shows weak oscillations near the graphene edge, revealing plasmons with a wavelength $\lambda_p$ of about 650 nm. We explain the existence of plasmons at the charge neutrality point by electron and hole populations that are thermally excited at room temperature[26, 27]. Their energy of about 25 meV is large enough for supporting GPs at THz frequencies (3.11 THz = 13 meV). In Fig. 4c we compare experimental (symbols, extracted from Fig. 4a) and calculated (solid line, see Methods) plasmon wavelengths. The calculation of the plasmon dispersion is carried out by considering the conductivity of graphene at finite temperature[28]. The excellent agreement verifies both the electrical tunability of acoustic GPs, as well as their existence at the charge neutrality point.

To study the acoustic plasmon amplitude decay time $\tau_p$ as a function of the charge carrier density $n_2$, we measure the decay length $L_{PC}$ of the photocurrent modulations in Fig. 4a by fitting a function that assumes both plasmon damping and radial-wave geometrical spreading (supplementary information S4). While such fitting is consistent with former near-field microscopy studies of GPs[29], $L_{PC}$ provides only qualitative estimates for the GP propagation lengths $L_P$, owing to the more complex GP detection mechanism based on the heating of the pn-junction both by propagating GPs and via heat transfer from the hot spot below the tip[14]. However, as $L_{PC}$ scales with $L_P$, the quantity $L_{PC}/v_g$ allows for studying how $\tau_p$ depends on the carrier density. Note that in our work low-energy plasmons are studied, thus allowing for analyzing plasmon damping at very low carrier densities, down to the charge neutrality point. We find that $L_{PC}/v_g$ increases with positive $n_2$, from about 380 fs for $n_2 = 1.9 \times 10^{11} cm^{-2}$ to about 600 fs for $n_2 = 7.7 \times 10^{11}$ cm$^{-2}$ (symbols in Fig. 4d). For negative $n_2$, $\tau_p$ is rather constant with a value of about 500 fs. These observations are consistent with the expected competition of two types of plasmon damping mechanisms: charge impurity and acoustic phonon scattering processes. For acoustic phonon scattering, a rather weak dependence of the GP lifetime on the carrier density is expected[30], which has been experimentally demonstrated for mid-infrared plasmons at high carrier densities[29]. However, for charge impurity scattering, a rather strong dependence of the GP lifetime on carrier density has been predicted[31]. Thus, for low carrier densities a cross-over is expected where charge impurity scattering dominates acoustic phonon scattering. We can qualitatively match the dependency of $L_{PC}/v_g$ on carrier concentration by microscopic plasmon lifetime calculations[11] (solid lines in Fig. 4d), which fully take into account the layers of our heterostructure (see supplementary information S8). Importantly, in these calculations we only consider Coulomb impurities[31] with a density $n_{\mathrm{imp}} = 7 \times 10^{10}$ cm$^{-2}$, thus supporting that long-range scattering agents play the dominant role for plasmon damping at low carrier densities. This is in strong contrast to plasmon damping in encapsulated graphene samples at high carrier densities, which is dominated by intrinsic acoustic phonons rather than by impurities[29].

The long-lived and strongly enhanced and confined fields of acoustic GPs could play an important role in fundamental studies of strong light-matter interactions at the nanoscale. Besides, their acoustic dispersion could offer manifold possibilities for the development of devices for detector, sensor and communication applications in the technologically important THz range, such as nanoscale waveguides or modulators. We also highlight that electrical detection of GPs constitutes an important technological advance in the field of graphene plasmonics, as purely on-chip functionalities can be now envisioned and developed. We finally stress that our imaging technique also opens the door to study and map local THz photocurrent in nanoscale semiconductor devices or 2D materials with unprecedented detail.

**Acknowledgements**

We thank Carlos Crespo for technical assistance with the THz laser. R.H., P.A-G., and A.N. acknowledge support from the Spanish Ministry of Economy and Competitiveness (national projects MAT2015-65525-R, FIS2014-60195-JIN, and MAT2014-53432-C5-4-R, respectively). F.K. acknowledges support by Fundacio Cellex Barcelona, the ERC Career integration grant (294056, GRANOP), the ERC starting grant (307806, CarbonLight), the Government of Catalonia trough the SGR grant (2014-SGR-1535), the Mineco grants Ramón y Cajal (RYC-2012-12281) and Plan Nacional (FIS2013-47161-P), and project GRASP (FP7-ICT-2013-613024-GRASP). R.H., F.K., A.P., and M.P. acknowledge support by the EC under the Graphene Flagship (contract no. CNECT-ICT-604391). Y.G. and J.H. acknowledge support from the US Office of Naval Research N00014-13-1- 0662.

**Author contributions**

P.A-G., A.J.H. and R.H. implemented the THz photocurrent near-field microscope. P.A-G. and A.W. performed the photocurrent nanoscopy experiments. A.Y.N. and A.W. performed the simulations of the GP near-field distributions and GP dispersion. Y.G. fabricated the h-BN/graphene/h-BN photodetector devices. A.P., N.F., and M.P. developed the analytic model for the acoustic GP dispersion and the theory of plasmon damping. W.Y. and S.V. fabricated photocurrent devices based on exfoliated and CVD grown graphene. A.J.H. contributed to the implementation of the THz photocurrent near-field microscope. K.W. and T.T. synthesized the h-BN. F.C., L.E.H., and J.H. coordinated the fabrication of the different photocurrent samples. P.A-G., A.Y.N., A.W., M.B.L., M.P., F.H.L.M, and R.H. analyzed the data. P.A-G., A.Y.N, and R.H. wrote the manuscript with input from all authors. All authors contributed to the scientific discussion and manuscript revisions.

**Additional information**

Supplementary information is available in the online version of the paper. Reprints and permissions information is available online at www.nature.com/reprints. Correspondence and requests for materials should be addressed to P.A.-G. and R.H.

**Competing financial interests**

R.H. is co-founder of Neaspec GmbH, a company producing scattering-type scanning near-field optical microscope systems such as the one used in this study. All other authors declare no competing financial interests.

**Methods**

*Determination of local carrier densities.*

The near-field photocurrent profiles shown in Fig. 2a were extracted at a distance of about 1 μm from the 50 nm-wide gap between the two gates, in order to ensure a well-defined and homogeneous carrier density in the graphene sheet. As can be seen in the supplementary information S4.2, the period of the signal oscillations (i.e. the GP wavelength) is constant for all near-field photocurrent line profiles taken at distances d between 250 nm and 1500 nm from the gap. Because the GP wavelength depends on the carrier concentration, we can conclude that the carrier concentration is well established at d = 1 μm.

The sheet carrier densities $n_{L,R}$ were obtained from the gate voltages $U_{L,R}$ according to $n_{L,R} = k(U_{L,R} - U_{off}) = (0.464 \times 10^{16} \text{ m}^{-2} \text{ V}^{-1})(U_{L,R} - 0.34 \text{ V})$. The offset voltage $U_{off}$ corresponds to the gate voltage for which the resistance of the graphene sheet is maximum. The coefficient $k$ is the calculated electrostatic capacitance of the 42 nm h-BN layer, assuming dielectric constant of 3.56 for h-BN[32, 33].

*Calculation of acoustic GP dispersion*

The graphene plasmon modes in both the thin-film stack of vacuum–SiO$_2$(285 nm)–AuPd(10nm)-h-BN(42nm)–graphene–h-BN(13 nm)–vacuum and free-standing graphene (air-graphene-air) were calculated using the electromagnetic transfer matrix method. The finite-temperature local random phase approximation[28] was used to calculate the graphene conductivity σ(ω) at room temperature. We assume high-quality graphene with a mobility of 40000 cm$^2$/V·s (ref. [15]). The permittivity model used for the h-BN films was the one of ref.[29]. The imaginary part of the Fresnel reflection coefficient is displayed as blue color plot in Fig. 2b.

*Numerical calculations of spatial near-field distributions*

Finite-elements numerical simulations (Comsol software) were used to calculate the spatial distribution of the vertical near-field component in heterostructure and around the free-standing graphene sheet, both shown in Fig. 3. The conductivity of graphene was calculated according to the local random phase approximation[28]. We assume high-quality graphene with a mobility of 40000 cm$^2$/V·s (ref. [15]). Further, we assumed a spatially constant carrier concentration (for justification see the Methods section *Determination of local carrier densities*). We also excluded reflection of GPs at the pn-junction, which we justify by the absence of photocurrent oscillations with half the GP wavelength perpendicular to the pn-junction. Because of the absence of GP reflections, the pn-junction essentially damps or transmits GPs. Altogether, we can conclude that the GP propagation parallel to the gap, and at 1 μm distance, is not affected by the pn-junction. In our numerical calculations we can thus model our experiment by a planar heterostructure with spatially uniform carrier concentration.

**Figures and captions**

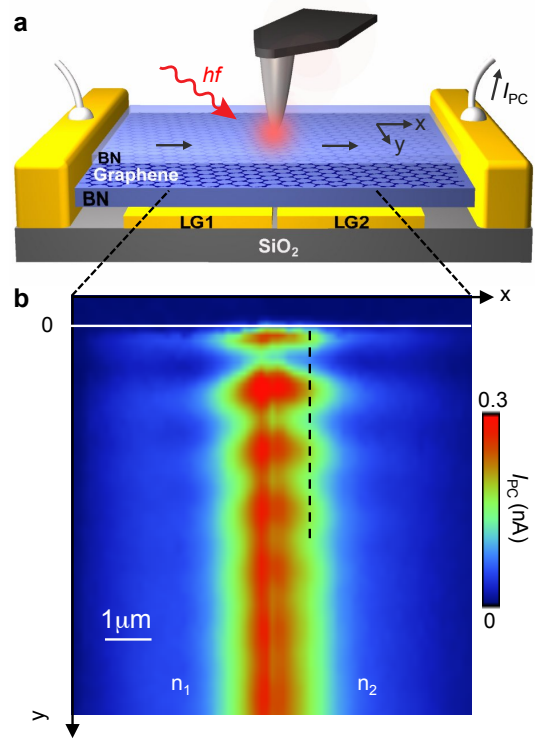

**Fig. 1. THz photocurrent nanoscopy of graphene plasmons in a split-gate photodetector.** (**a**) Schematics of the experimental setup. The laser-illuminated metal tip of an atomic force microscope (AFM) serves as a nanoscale near-field light source. The near-field induced photocurrent in the graphene (encapsulated by h-BN layers) is measured through the two metal contacts to the left and right. LG1 and LG2 represent the split gate (gold) used for controlling the carrier concentration in the graphene to the left and the right of the gap between them. (**b**) Experimental near-field photocurrent image, $I_{PC}$, recorded at $f = 2.52$ THz. The carrier densities were chosen to be $n_1 = 0.77 \times 10^{12}$ cm$^{-2}$ and $n_2 = -0.77 \times 10^{12}$ cm$^{-2}$. The horizontal white solid line marks the edge of the graphene sheet.

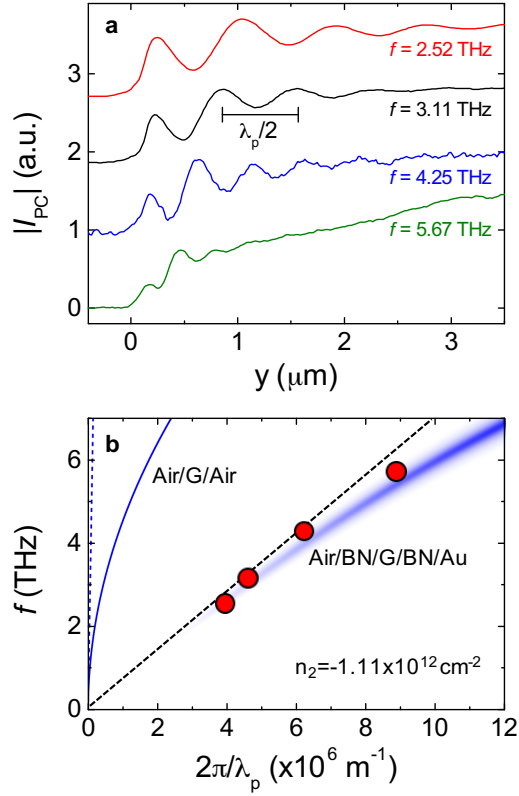

**Fig. 2. THz graphene plasmon wavelengths and dispersion.** (**a**) Near-field photocurrent profiles at different illumination frequencies *f*. They were recorded along the black dashed line in Fig. 1b. The peak-to-peak distance reveals $\lambda_p/2$. The carrier density was $n_2 = -1.11 \times 10^{12}$ cm$^{-2}$. (**b**) Experimental and theoretical dispersion relation. The red symbols display the experimental values extracted from Fig. 2a. Error bars (standard deviation) are within the symbols size (see supplementary information S4). The blue color plot shows the calculated dispersion of graphene plasmons in a Air/BN/graphene/BN/AuPd/SiO$_2$ heterostructure assuming the experimental carrier concentration and layer thicknesses (see Methods). The thin blue solid line shows the calculated plasmon dispersion for free-standing graphene (Air/G/Air) of the same carrier concentration. The dashed black line displays the plasmon dispersion calculated according to Eq. 1. The dashed blue line indicates the light line in free-space.

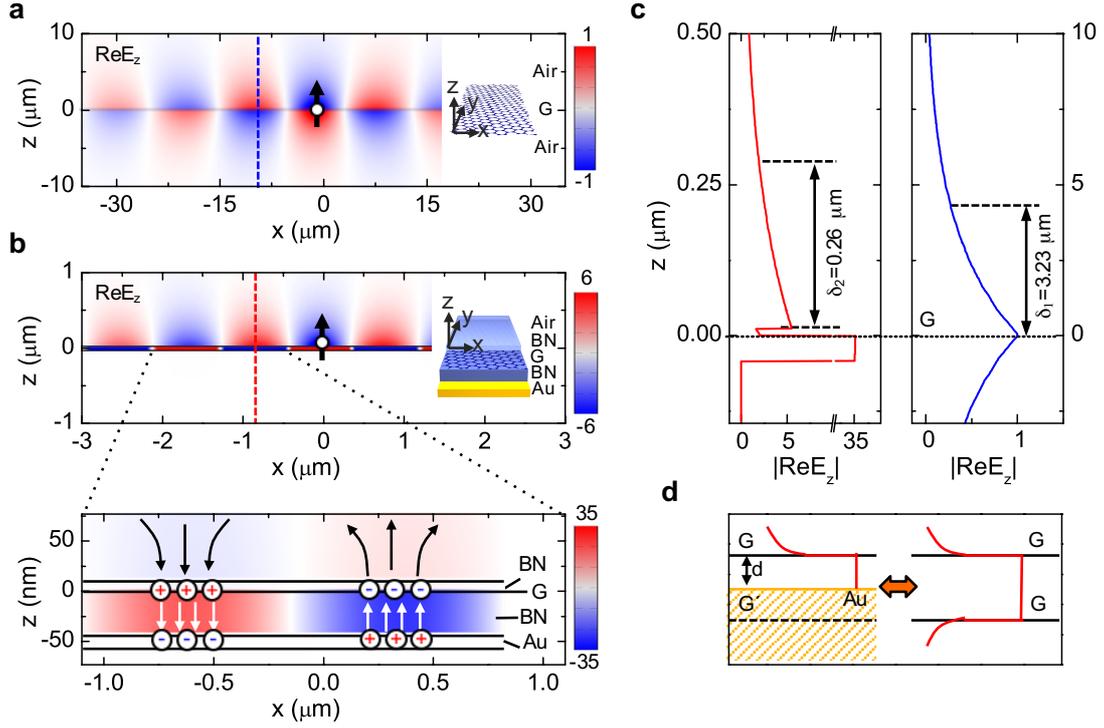

**Fig. 3. Near-field distribution of THz graphene plasmons.** (**a,b**) Numerical simulations of the near-field distribution of THz graphene plasmons excited by a point dipole source located above (**a**) a free-standing graphene sheet (Air/G/Air) and (**b**) an air/BN/G/BN/AuPd/SiO$_2$ heterostructure assuming the experimental layer thicknesses. The real part of the vertical field component, Re(E$_z$(x,z)) at a frequency of 2.52 THz is shown for both cases. The + and – symbols in the zoom-in image in (b) sketch the charge distribution in graphene and AuPd. (**c**) Near-field profiles |Re(E$_z$)| perpendicular to the graphene surface. Left: profile along the dashed red line in (b). Right : profile along the dashed blue line (a). Both profiles were normalized to the maximum of |Re(E$_z$)| on top of a free-standing graphene sheet. (**d**) Schematics of the plasmonic near-field profile for a graphene sheet above a gold surface (left) and for two parallel graphene sheets. The distance between the two graphene sheets is twice the distance between graphene and gold surface.

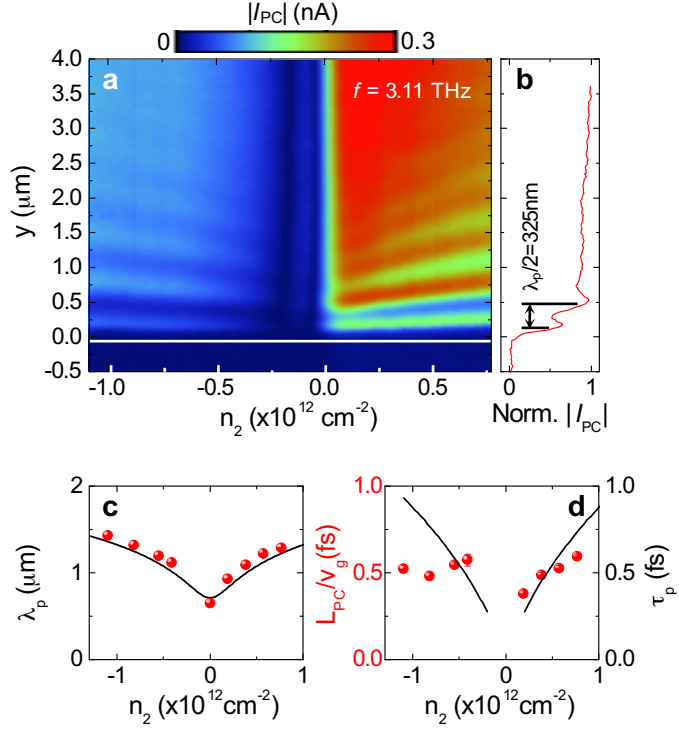

**Fig. 4. Wavelength and amplitude decay time of acoustic THz graphene plasmons as a function of carrier density.** (**a**) Near-field photocurrent signal $|I_{PC}|$ as a function of tip position y (perpendicular to the graphene edge marked by horizontal solid white line) and carrier density $n_2$, recorded at an illumination frequency $f$ = 3.11 THz and $n_1$ = -1.4x10$^{11}$ cm$^{-2}$. For understanding of the observed two $|I_{PC}|$ minima see supplementary information S2. (**b**) Photocurrent profile at the charge neutrality point (along the dotted vertical while line in (**a**)). (**c**) Experimental (symbols, extracted from Fig. 4a) and calculated (black solid line) plasmon wavelengths $\lambda_p$. (**d**) $L_{PC}/v_g$ as a function of carrier density $n_2$. Symbols display experimental values extracted from (**a**). The black solid line shows the calculated plasmon lifetime $\tau_p$, where the density of Coulomb impurities was used as a fitting parameter.